Quantum – coherent dynamics in photosynthetic charge separation revealed by wavelet analysis

*Elisabet Romero*[*†], *Javier Prior*[*‡], *Alex W. Chin*[§], *Sarah E. Morgan*[§], *Vladimir I. Novoderezhkin*[#], *Martin B. Plenio*[⊥], *and Rienk van Grondelle*[†]

[†]Department of Physics and Astronomy, VU University Amsterdam, The Netherlands
[‡]Departamento de física aplicada, Universidad Politécnica de Cartagena, Spain
[§]Cavendish Laboratory, University of Cambridge, United Kingdom
[#]A. N. Belozersky Institute of Physico-Chemical Biology, Moscow State University, Russia
[⊥]Institute of Theoretical Physics and IQST, Universität Ulm, Germany

AUTHOR INFORMATION

**Corresponding Authors**
*Elisabet Romero, De Boelelaan 1081, 1081 HV, Amsterdam, The Netherlands, e-mail: eli@few.vu.nl;
*Javier Prior, Paseo Alfonso XIII, 30203 Cartagena, Spain, e-mail: Javier.Prior@upct.es.
These authors contributed equally to this work.

ABSTRACT

Experimental/theoretical evidence for sustained vibration-assisted electronic (vibronic) coherence in the Photosystem II Reaction Center (PSII RC) indicates that photosynthetic solar-energy conversion might be optimized through the interplay of electronic and vibrational quantum dynamics. This evidence has been obtained by investigating the primary charge separation process in the PSII RC by two-dimensional electronic spectroscopy (2DES) and Redfield modeling of the experimental data. However, while conventional Fourier transform analysis of the 2DES data allows oscillatory signatures of vibronic coherence to be identified in the frequency domain in the form of *static* 2D frequency maps, the real-time evolution of the coherences is lost. Here we apply for the first time wavelet analysis to the PSII RC 2DES data to obtain *time-resolved* 2D frequency maps. These maps allow us to demonstrate that i) coherence between the excitons initiating the two different charge separation pathways is active for more than 500 fs, and ii) coherence between exciton and charge-transfer states, the reactant and product of the charge separation reaction, respectively; is active for at least 1 ps. These findings imply that the PSII RC employs coherence i) to sample competing electron transfer pathways, and ii) to perform directed, ultrafast and efficient electron transfer.

**Introduction**

In oxygenic photosynthesis the site of solar-energy conversion, the photosystem II reaction center (PSII RC), is a membrane-bound pigment-protein complex that converts solar photons into a stable charge-separated state which, in turn, creates an electrochemical potential across the membrane. Remarkably, charge separation in the PSII RC takes place with near-unity quantum efficiency despite the intrinsically highly disordered energy landscape of this complex[1-2]. The energetic disorder arises from fast nuclear motions of all atoms in the system (dynamic disorder) as well as from slow collective protein motions (static disorder). The static disorder plays a central role in the PSII RC dynamics because it generates a variety of different protein conformations within the sample ensemble, that is, a collection of energetically different RC complexes. From all possible realizations of the disorder (protein conformations) one may expect realizations optimal for ultrafast and efficient charge separation as well as non-optimal realizations that may lead to unwanted energy losses. Nevertheless, the near-unity charge separation quantum efficiency in the PSII RC implies that the system is able to overcome energetic disorder.

The PSII RC contains four chlorophyll *a* (Chl) and two pheophytin *a* (Phe) molecules arranged in two quasi-symmetric branches ($D_1$ and $D_2$) in the center of the complex as well as two chlorophylls and two *β*-carotenes located at its periphery[3-4] (Fig. 1). Upon excitation, several delocalized and collective exciton states with significant charge-transfer (CT) character are created[5-7] (in other words, mixed exciton-CT states). In this scenario, charge separation proceeds via two different routes, the $Chl_{D1}$ and $P_{D1}$ pathways[8-9], which provide functional flexibility to the complex. The two pathways are:

$Chl_{D1}$ path: 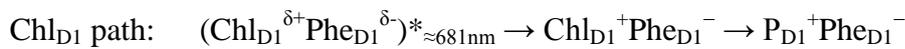 $(Chl_{D1}^{\delta+}Phe_{D1}^{\delta-})^*_{\approx 681nm} \rightarrow Chl_{D1}^+Phe_{D1}^- \rightarrow P_{D1}^+Phe_{D1}^-$

$P_{D1}$ path: 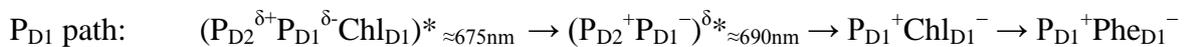 $(P_{D2}^{\delta+}P_{D1}^{\delta-}Chl_{D1})^*_{\approx 675nm} \rightarrow (P_{D2}^+P_{D1}^-)^{\delta*}_{\approx 690nm} \rightarrow P_{D1}^+Chl_{D1}^- \rightarrow P_{D1}^+Phe_{D1}^-$

The energetically lowest exciton-CT states able to initiate charge separation are: $(Chl_{D1}^{\delta+}Phe_{D1}^{\delta-})^*_{\approx 681nm}$ and $(P_{D2}^{\delta+}P_{D1}^{\delta-}Chl_{D1})^*_{\approx 675nm}/(P_{D2}^+P_{D1}^-)^{\delta*}_{\approx 690nm}$, via the $Chl_{D1}$ and $P_{D1}$ paths, respectively[7-10] (Fig. 1) (the subscripts indicate the central absorption wavelength whereas the superscripts δ+/δ- and δ* indicate CT and exciton character, respectively). Since the participation of $Chl_{D1}$ in $(P_{D2}^{\delta+}P_{D1}^{\delta-}Chl_{D1})^*_{\approx 675nm}$ is small, in the following we refer to $(P_{D2}^{\delta+}P_{D1}^{\delta-})^*_{\approx 675nm}$ for simplicity. Note that $(P_{D2}^{\delta+}P_{D1}^{\delta-})^*_{\approx 675nm}$ has a high $(P_{D2}^{\delta+}P_{D1}^{\delta-})_+^*_{\approx 660nm}$ and a low $(P_{D2}^{\delta+}P_{D1}^{\delta-})_-^*_{\approx 675nm}$ energy component and that $(P_{D2}^+P_{D1}^-)^{\delta*}_{\approx 690nm}$ is a CT state with exciton character (CT-exciton)[7] (Fig. 1). Similarly, $(Chl_{D1}^{\delta+}Phe_{D1}^{\delta-})_-^*{}_{\approx 681nm}$ has a high energy counterpart which is difficult to observe spectroscopically due to spectral congestion but that in our model absorbs around 672 nm [that is: $(Chl_{D1}^{\delta+}Phe_{D1}^{\delta-})_+^*{}_{\approx 672nm}$][11].

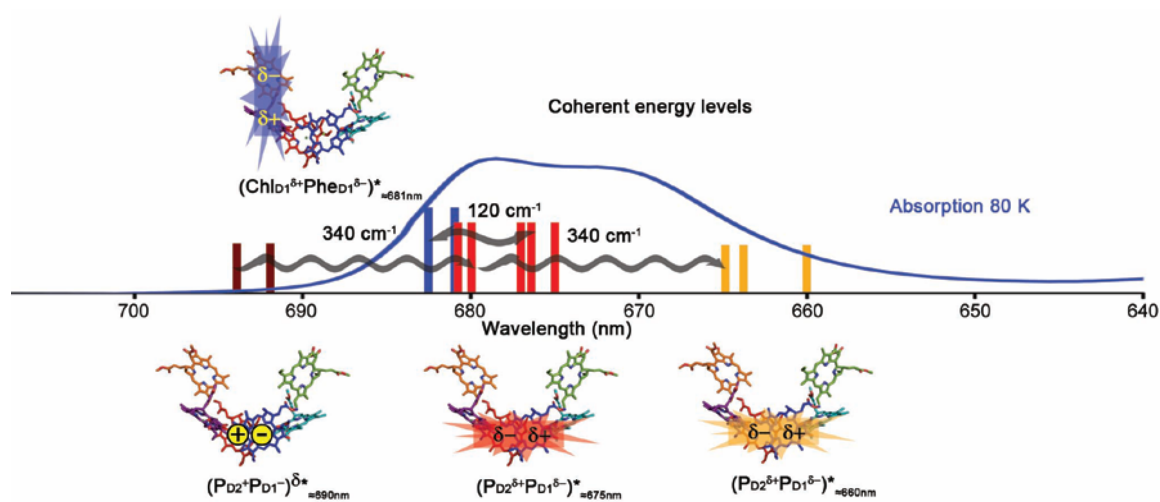

**Figure 1. PSII RC X-ray structure, electronic states involved in charge separation and absorption spectrum at 80 K.** The energy of the electronic states capable of initiating charge separation are shown as vertical lines on the absorption spectrum at 80 K (blue line) (the different lines for each state correspond to a different realization of the disorder). Only the coherences discussed in the text are shown for clarity. The cofactor participation in each electronic state and its electronic distribution is shown on top of the X-ray crystal structure of the PSII RC (cofactor arrangement adapted from ref. 3): stars and rectangles represent exciton and CT character, respectively. Color code: $(P_{D2}^{\delta+}P_{D1}^{\delta-})_+^*{}_{\approx 660nm}$ (orange), $(P_{D2}^{\delta+}P_{D1}^{\delta-})_-^*{}_{\approx 675nm}$ (red), $(P_{D2}^{+}P_{D1}^{-})^{\delta*}{}_{\approx 690nm}$ (dark red) and $(Chl_{D1}^{\delta+}Phe_{D1}^{\delta-})_-^*{}_{\approx 681nm}$ (blue). Cofactors colour code: $P_{D1}$ (red), $P_{D2}$ (blue), $Chl_{D1}$ (purple), $Chl_{D2}$ (cyan), $Phe_{D1}$ (orange), $Phe_{D2}$ (green). The horizontal wiggled arrows represent the discussed coherences between electronic states observed as cross-peaks in the 120 and 340 cm$^{-1}$ time-resolved 2D frequency maps.

Recently we have demonstrated that vibration-assisted electronic (vibronic) coherence is at play in the PSII RC charge separation process at physiological temperature[11-13]. More specifically, electronic coherence between the above mentioned electronic states in the system is maintained by the coupling of Chl *a* intra-molecular vibrational modes to these states (provided that the vibrational modes match the energy gap between the states) (Fig. 1). This vibronic mechanism has been proposed and described by theoretical methods[14-21] and later demonstrated by combined experimental-theoretical approaches[11-13,22-24]. In our previous work[12], two-dimensional electronic spectroscopy (2DES)[25-26] and Fourier transform (FT) analysis allowed to visualize the coherence between electronic states in 2D frequency maps [obtained by Fourier transformation of the real rephasing 2D spectra along the population time ($T$)]. These maps display features corresponding only to the states oscillating at a certain frequency and, therefore, are more selective than the 2D spectra. Moreover, two main 2D frequency maps characteristics: i) the oscillation frequency (in or out of resonance with the electronic states energy gaps), and ii) the position of the most intense bands (corresponding or not to the states in resonance with the oscillation frequency); provide an invaluable tool to classify the observed coherences as mainly electronic, mixed electronic-vibrational (vibronic) or mainly vibrational[11,13,27-28]. We found that three main frequency modes, 120, 340 and 730 cm$^{-1}$, representative of mainly electronic, mixed electronic-vibrational (vibronic) and mainly vibrational coherence, respectively, dominate the oscillatory dynamics[11,13]. The first two types of coherences have functional significance and will be analyzed and discussed below, the non-functional vibrational coherence will be analyzed and discussed in the Supplementary Information (Fig. S1-S2). The 120 and 340 cm$^{-1}$ frequency modes couple to the states involved in charge separation in the PSII RC and allow the complex to rapidly and coherently sample its energy landscape. The 120 cm$^{-1}$ mode coherently mixes the two lowest exciton-CT states capable of initiating charge separation [$(P_{D2}{}^{\delta+}P_{D1}{}^{\delta-})^*_{\approx 675nm}$ and $(Chl_{D1}{}^{\delta+}Phe_{D1}{}^{\delta-})^*_{\approx 681nm}$] (Fig. 1) via the $P_{D1}$ and $Chl_{D1}$ pathways, respectively, and therefore, upon excitation it creates and maintains the coherence between them. The most crucial consequence of this coupling is that it allows the complex to sample the $Chl_{D1}$ and $P_{D1}$ pathways, and therefore, to select the most optimal charge-separation pathway for a specific realization of the disorder. The 340 cm$^{-1}$ mode couples to the three states involved in the $P_{D1}$ pathway [$(P_{D2}{}^{\delta+}P_{D1}{}^{\delta-})_+{}^*_{\approx 660nm}$, $(P_{D2}{}^{\delta+}P_{D1}{}^{\delta-})_-{}^*_{\approx 675nm}$, and $(P_{D2}{}^{+}P_{D1}{}^{-})^{\delta*}_{\approx 690nm}$] (Fig. 1). This mode promotes ultrafast exciton relaxation from $(P_{D2}{}^{\delta+}P_{D1}{}^{\delta-})_+{}^*_{\approx 660nm}$ to $(P_{D2}{}^{\delta+}P_{D1}{}^{\delta-})_-{}^*_{\approx 675nm}$ and, most importantly, it allows coherent electron transfer from $(P_{D2}{}^{\delta+}P_{D1}{}^{\delta-})_-{}^*_{\approx 675nm}$ to $(P_{D2}{}^+P_{D1}{}^-)^{\delta}*_{\approx 690nm}$, strongly suggesting that charge separation in the PSII RC proceeds via a coherent mechanism[12].

However, even though the 2D frequency maps provide crucial information about coherences between electronic states, these maps are a static representation due to the fact that the time information has been exchanged by frequency information after FT of the 2D spectral traces along $T$. For that reason, here we investigate the dynamics of coherence during the process of charge separation in the PSII RC by applying a novel analysis of 2DES data whose advantages have only very recently been demonstrated[29-30]: Wavelet Transform (WT)[31-32]. The WT technique is a powerful tool to investigate coherent dynamics in multichromophoric systems due to the fact that it combines the frequency and time information available from 2DES.

At this point, it is important to emphasize that the 2DES[25-26] experiment is a tool to visualize the existence of strongly coupled pigment chains by the coherent excitation of the corresponding electronic states. If any pair of states contains a mixing of pigments, such as the exciton-CT states presented above, their coherent excitation will generate dynamic coherence visualized as amplitude oscillations in the 2D traces. In this manner the dynamic coherence between the states created by the coherent broadband and ultrashort laser excitation employed in the 2DES experiment highlights the inner coherence (that is the coherent mixing of the pigments within states)[11,13]. It is important to realize that the inner coherence is an intrinsic property of the system, and therefore, its presence is consistent with directed and efficient energy and electron transfer also upon non-coherent excitation, specifically in natural photosynthesis under sunlight illumination.

**Results**

*Wavelet transform analysis*

The continuous wavelet transform (CWT) analysis is applied to the PSII RC 2DES dataset to unravel the dynamics of the coherences previously resolved by Fourier transform analysis (the details are provided in the *Methods* section). Briefly, CWT allows to extract both the frequency and time information contained in 2DES experimental data, however, a compromise between frequency and time resolution must be found; extremely high frequency resolution will result in poor time resolution and vice versa [a comparison of two limiting cases together with the optimal case used in the analysis presented here can be found in the Supplementary Information (Fig. S3)]. We perform CWT analysis for each 2D spectral trace in order to obtain *time-resolved* 2D frequency maps. In the following we analyze the time evolution of the most representative features in the 120 and 340 cm$^{-1}$ 2D frequency maps at 80K, that is, the features corresponding to coherence between the electronic states involved in charge separation (the CWT analysis for the diagonal bands is shown and discussed in the Supplementary Information, Fig. S4). In our previous work both the 80 K and room temperature (277 K) 2D experimental data was presented and it was concluded that both data sets reflect the same coherences. Since the 80 K data set has an enhanced spectral resolution, it is presented here whereas the wavelet analysis of the room temperature dataset is displayed in the Supplemental Information (Fig. S2 for the 730 cm$^{-1}$ and Fig. S5 for the 120 and 340 cm$^{-1}$ vibrational modes).

The 120 ± 20 cm$^{-1}$ *time-resolved* 2D frequency maps at 80 K obtained by CWT are shown in Fig. 2a together with the dynamics for the ($\lambda_\tau$, $\lambda_t$) equal to (675, 681) and (677, 683) nm cross-peaks (2D wavelet traces) (Fig. 2b). The range ± 20 cm$^{-1}$ indicates the frequency resolution of the experimental 2D data taking into account the 20 fs step in population time (*T*) and the *T* range (80-1000 fs) used to apply the FT and the wavelet analysis. The amplitude of these cross-peaks correspond to the coherence between $(P_{D2}^{\delta+}P_{D1}^{\delta-})^*_{\approx 675\text{-}677nm}$ and $(Chl_{D1}^{\delta+}Phe_{D1}^{\delta-})^*_{\approx 681\text{-}683nm}$ exciton-CT states. Note that the two distinguishable maxima indicate the presence of two groups of realizations of the disorder with the central absorption wavelength indicated above. However, it should be noted that other realizations where the $(P_{D2}^{\delta+}P_{D1}^{\delta-})^*$ and $(Chl_{D1}^{\delta+}Phe_{D1}^{\delta-})^*$ states absorb at higher and lower energies are also present [for a detailed explanation on this topic see the Supplementary Information (Fig. S6)]. The 2D wavelet traces in Fig. 2b show that the coherence between these two states is sizeable for more than 500 fs [with a decay time of 520 fs (1/e)]. Similarly, the room temperature data shows a decay time of 660 fs (1/e) (Fig. S5b). According to the vibronic mechanism presented by several theoretical groups[14-21] and confirmed by experimental and/or combined experimental/theoretical approaches[11-13,22-24], the above mentioned observation indicates that the 120 cm$^{-1}$ intra-molecular vibrational mode maintains the electronic coherence between these two states during the time scale of charge separation[12,33-39]. Remarkably, since these states are the precursors of the $P_{D1}$ and $Chl_{D1}$ charge separation pathways active in the PSII RC, the long-lived coherence between them allows the system to simultaneously access these two states in order to select the most optimal charge separation pathway depending on the specific realization of the disorder at both cryogenic (80 K) and room temperature (277 K). The coherence dynamics unravelled by CWT analysis provides two relevant pieces of information: i) the speed of coherent charge separation: the decay rate indicates how fast the RCs perform charge separation, given that the coherence is lost once the first charge-separated state is formed, and ii) the time range for which the RCs are able to maintain coherence between the states which initiate charge separation.

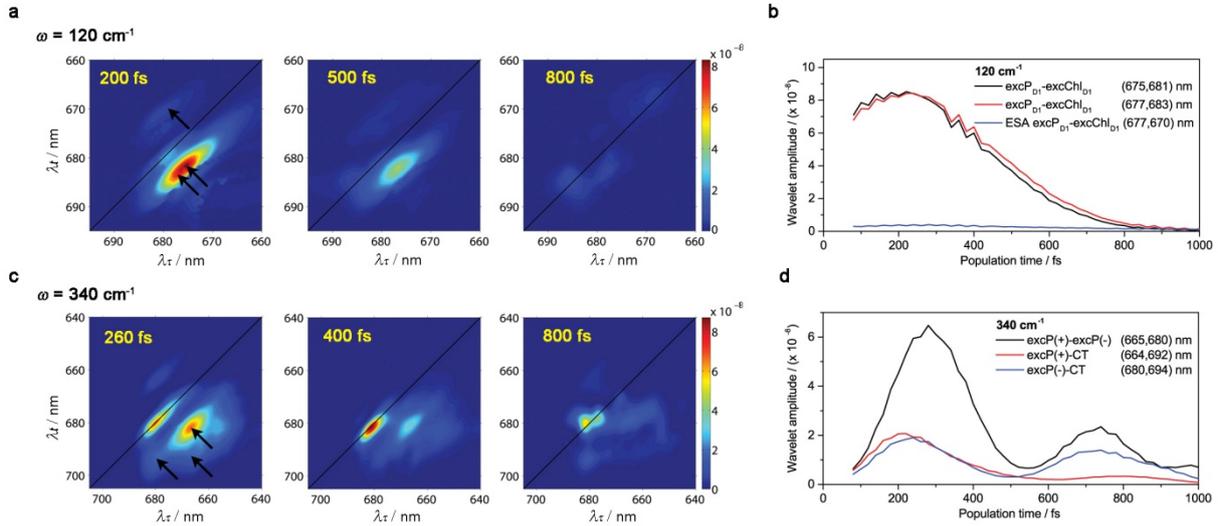

**Figure 2. Dynamics of vibronic coherence in the PSII RC at 80K: time-resolved 2D frequency maps (a,c) and wavelet traces (b,d).** (**a**) 120 cm$^{-1}$ time-resolved 2D frequency maps at $T$ = 200, 500 and 800 fs. The black arrows in the 200 fs time-resolved 2D frequency map indicate the position of the selected 2D traces. (**b**) 120 cm$^{-1}$ 2D wavelet traces. (**c**) 340 cm$^{-1}$ time-resolved 2D frequency maps at $T$ = 260, 400 and 800 fs. The black arrows in the 260 fs time-resolved 2D frequency map indicate the position of the selected 2D wavelet traces. (**d**) 340 cm$^{-1}$ 2D wavelet traces. The black arrows in the 200 fs time-resolved 2D frequency map indicate the position of the selected 2D wavelet traces.

The 340 ± 20 cm$^{-1}$ *time-resolved* 2D frequency maps at 80K obtained by CWT are shown in Fig. 2c together with the 2D wavelet traces for the (665, 680) nm cross-peak [coherence between $(P_{D2}^{\delta+}P_{D1}^{\delta-})_{+}*_{\approx 660nm}$ and $(P_{D2}^{\delta+}P_{D1}^{\delta-})_{-}*_{\approx 675nm}$], and the (680, 694) nm cross-peak [coherence between $(P_{D2}^{\delta+}P_{D1}^{\delta-})_{-}*_{\approx 675nm}$ and $(P_{D2}^{+}P_{D1}^{-})^{\delta}*_{\approx 690nm}$]$^{12}$ (Fig. 2d). The cross-peak traces show a rise in the 80-280 fs range, a decay in the 280-500 fs range and a second rise and decay in the 500-700 fs and 700-1000 fs ranges, respectively. The room temperature 2D wavelet traces shown in Fig. S5d exhibit very similar dynamics. The observed rise and decay of the wavelet amplitudes may contain information about coherent dynamics, as they may arise from reversible coherence transfer, coherence-to-population transfer or population-to-coherence transfer between the electronic states involved in the $P_{D1}$ charge separation pathway. However, due to the close proximity of the observed beating frequencies in the PSII RC, these rise and decay dynamics may be influenced by the presence of interferences between these frequencies. If this is the case, the amplitude of the 340 cm$^{-1}$ feature should oscillate with a period inversely proportional to the frequency spacing between the interfering frequencies. In order to test this hypothesis, we have measured the period of the rise/decay features and calculated the group of frequencies that could give rise to such interference artefact. The measured periods for the 2D wavelet traces shown in Fig. 2d are: ≈ 500 fs (≈ 70 cm$^{-1}$) for both the (665, 680) nm and (680, 694) nm cross-peaks. Therefore, the obtained interfering frequencies for the below diagonal cross-peaks are: 340 – 70 = 270 cm$^{-1}$ and 340 + 70 = 410 cm$^{-1}$. Not surprisingly, the calculated interfering frequencies, that is: 270 and 410 cm$^{-1}$, are close to some of the closely spaced beating frequencies experimentally observed and present in the 2D spectral data[12]: 265 and 440 cm$^{-1}$. This correspondence indicates that, at least in part, the observed dynamics may arise from interference effects rather than entirely from vibronic dynamics.

*Windowed Fourier transform*

In order to test the effect of closely spaced interfering frequencies in the dynamics of the coherences present in the 2D data, we have performed a windowed Fourier transform analysis of the (665,680) nm 2D trace in the 340 cm$^{-1}$ 2D frequency map at 80 K (Supplementary Information, Fig. S7). Three windows centred at 340 cm$^{-1}$ with full width at half maximum of 20, 40 and 80 cm$^{-1}$ have been convoluted with the Fourier transform. The comparison of the original 2D trace with the windowed 2D trace clearly shows that only when closely spaced frequencies are included by the window, the interference artefact appears as an envelope with a 500 fs period (Fig. S7b and S7c). These results demonstrate that the two rise/decay features with a 500 fs period obtained by CWT do not correspond to coherent dynamics, they are the result of interference between 340 cm$^{-1}$ frequency and other closely spaced frequencies present in the 2D data. It is worth noting that the existence of several frequencies in the PSII RC 2D data is a consequence of the complexity of the system, more specifically, a consequence of the presence of several states probed simultaneously together with severe spectral congestion.

In relation to the CWT analysis, the windowed Fourier transform analysis shows that the interference artefact is not introduced by the CWT analysis, the interference artefact is a consequence of the compromise between frequency and time resolution that needs to be made when performing the CWT analysis (for details see the Methods section).

*Simulation of the 2D spectral dynamics*

In order to investigate further the effect of interfering frequencies in the coherent dynamics and aiming to differentiate between the *real* dynamics and the interference artefact, we have performed a simulation of the time evolution of the real rephasing 2D spectra for a single optical excited state linearly coupled to modes of frequency 120, 190, 265, 340 and 440 cm$^{-1}$ (that is, the frequencies present in the experimental PSII RC 2DES data[12]). The 2D signal was generated using the nonlinear response function formalism presented previously[40], assuming the impulsive limit for the laser excitation. The microscopic model consists of a single optical transition between ground and excited state with the excited state energy surface displaced relative to the ground state along the vibrational coordinates. An additional overdamped brownian oscillator coordinate has been included to account for the residual dephasing environment[41]. The influence of the intra-molecular vibrational modes and the protein environment on the 2D signal was accounted for by using the line shape approach[40-41]. The environment is characterized by a reorganization energy of 50 cm$^{-1}$ and a temperature of 80 K. All vibrational modes were coupled to the excited state with a Huang-Rhys factor equal to 0.1 (qualitatively similar results are obtained for smaller Huang-Rhys factors). By including both ground state bleach and stimulated emission (no decay of the excited state is assumed), the 3rd order response function can be written down analytically in the time domain[40], and the 340 cm$^{-1}$ 2D frequency map can be obtained by numerical Fourier transform (Fig. 3b).

Applying the CWT to the simulated real rephasing 2D spectra at the peaks (A-D) highlighted in Fig. 3b, we find that their dynamics (represented in scalograms, that is, frequency – time plots) (Fig. 3d) are dominated by a pseudofrequency around 340 cm$^{-1}$ which shows slow amplitude oscillations (period T ≈ 500 fs), as observed in the experimental PSII RC 2D spectral data (Fig. 2d and 3c). Note that the diagonal peak at (681,680) nm displays a more complex pattern most likely due to the population of several electronic states absorbing at 680 nm and the presence of the non-trivial interplay between mixed exciton-vibrational coherences[11,13] (for more details see the Supplementary Information, Fig. S4).

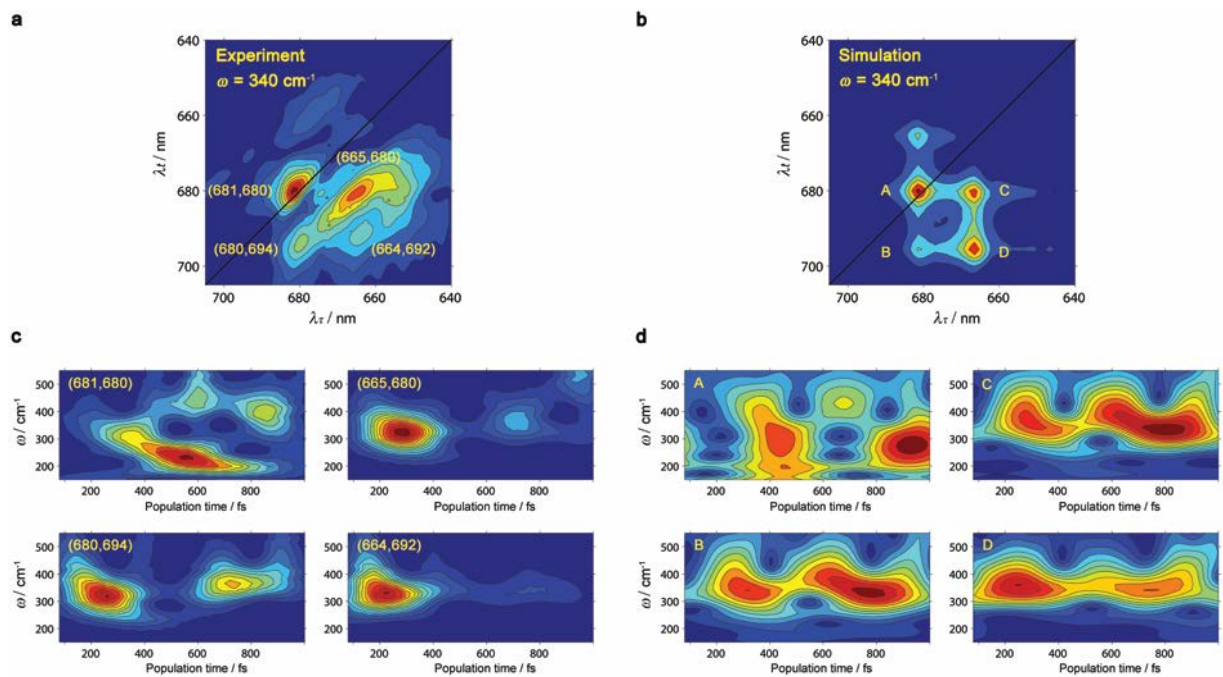

**Figure 3. Comparison of experimental vibronic coherence dynamics in PSII RC at 80 K and simulations of the classical undamped wave packet dynamics of a single electronic transition (simulation details given in the main text).** (**a**) The experimental 2D frequency map at 340 cm$^{-1}$. (**b**) The simulated 2D frequency map at 340 cm$^{-1}$, showing the predicted five peak pattern[11,13,17]. (**c**) Experimental wavelet scalograms computed at the peak positions indicated in (**a**). (**d**) Wavelet scalograms computed for the simulated peak positions indicated in (**b**). Recurrence of signals in both (**c**) and (**d**) are consistent with the beating envelope due to vibrational coherences with closely spaced frequencies, but the decay of signals in (**c**) points to additional electronic dynamics at certain peak positions.

As it has been shown above by the windowed Fourier transform analysis, the slow amplitude oscillations with period T ≈ 500 fs observed in all the calculated scalograms arise from interference between the 265 cm$^{-1}$ and 440 cm$^{-1}$ vibrational modes with the 340 cm$^{-1}$ mode (Supplementary Information Fig. S7 and Fig. 3c,d). Consequently, our simulation shows that the slow recurrences observed in the 2D wavelet traces (and in the scalograms) can be reproduced by interference between the closely spaced vibrational modes present in the PSII RC. Hence, we conclude that the rise/decay features with period T ≈ 500 fs arise mainly from interference effects (interference artefact) and not from reversible coherence transfer, coherence-to-population transfer or population-to-coherence transfer (*real* dynamics).

We also note that several other minor features in the theoretical scalograms are rather different from the experimentally observed wavelet scalograms. For example, the theoretical scalograms in Fig. 3b-c show an apparent oscillatory shift in pseudofrequency over the course of a picosecond that is absent in the experimental scalograms. As discussed above, beats (with periods matching frequency differences) appear due to the overlapping of several closely spaced frequencies, and the broad range of vibrational frequencies present means that many of such beating signals will appear in a typical scalogram. The oscillatory shifting is due in particular to the strong beating nodes that appear at the "top" and "bottom" edges of the scalogram at slightly different times (they appear to have roughly the same period but have a phase lag between them). The most likely cause for the differences between the theoretical data and the experimental signal is damping; the vibrational coherence is undamped in the simple theoretical signals, but coherences in the experimental data are likely to decay at varying rates over a picosecond. As a result of this, the experimental signal is dominated by a smaller band of pseudofrequencies (those with the smallest dephasing rates) at longer times and its scalogram shows less beating structure in its time evolution.

However, excluding the slow recurrences and the minor differences discussed above, there is a remarkable and clear difference between simulated and experimental dynamics: the simulated data shows no decay (since the underlying model does not include any electronic dynamics) whereas the experimental data shows a different decay pattern for the different cross-peaks highlighted in Fig. 3a (Fig. 3c). These different decay dynamics are also observed after applying the windowed Fourier transform to these cross-peaks (Fig. S8). Therefore, in spite of the unavoidable presence of interference effects, the overall dynamics of the coherences observed in the experimental *time-resolved* 2D frequency maps contains real information which we interpret and discuss in the *Discussion* section.

**Discussion**

On the one hand, the interpretation of the 120 cm$^{-1}$ *time-resolved* 2D frequency map at both 80 K and 277 K demonstrates that the coherence between the precursors of the two different charge separation pathways [$(P_{D2}^{\delta+}P_{D1}^{\delta-})^*_{\approx 675nm}$ and $(Chl_{D1}^{\delta+}Phe_{D1}^{\delta-})^*_{\approx 681nm}$ for the $P_{D1}$ and $Chl_{D1}$ pathways, respectively] is maintained in a significant fraction of PSII RC complexes for more than 500 fs following photo-excitation, allowing the system to coherently sample its energy landscape and therefore, to choose the most optimal pathway according to each specific realization of the disorder.

On the other hand, we have unequivocally shown that all the cross-peak features in the 340 cm$^{-1}$ *time-resolved* 2D frequency map obtained by CWT suffer from interference effects revealed by an amplitude beating with a 500 fs period by three independent methods: i) the calculation of the expected interfering frequencies based on the observed period of rise/decay features and the fact that those frequencies correspond to the experimentally observed frequencies, ii) the windowed Fourier transform analysis that shows that the rise/decay features are a consequence of interference between 340 cm$^{-1}$ frequency and other frequencies present in the 2D data, and iii) the simulation of the dynamics including the experimentally observed frequencies that display the same rise/decay features as the ones present after the CWT analysis. However, even in the presence of interference effects, the CWT analysis of the 2D experimental data at 80 K reveals a remarkable fact: the (665, 680) and (664, 692) nm cross-peaks [coherence between $(P_{D2}{}^{\delta+}P_{D1}{}^{\delta-})_+{}^*{}_{\approx 660nm}$ and $(P_{D2}{}^{\delta+}P_{D1}{}^{\delta-})_-{}^*{}_{\approx 675nm}$ and coherence between $(P_{D2}{}^{\delta+}P_{D1}{}^{\delta-})_+{}^*{}_{\approx 660nm}$ and $(P_{D2}{}^+P_{D1}{}^-)^{\delta*}{}_{\approx 690nm}$] display a considerable decay from 300 to 700 fs (3.3- and 6.7-fold decay, respectively) whereas the (680, 694) nm cross-peak [coherence between $(P_{D2}{}^{\delta+}P_{D1}{}^{\delta-})_-{}^*{}_{\approx 675nm}$ and $(P_{D2}{}^+P_{D1}{}^-)^{\delta*}{}_{\approx 690nm}$] shows a minor decay (1.4-fold decay) (Fig. 3 and S8) (similar dynamics are observed at room temperature, Fig. S5d). In other words, the coherence involving excitation of the high exciton component $(P_{D2}{}^{\delta+}P_{D1}{}^{\delta-})_+{}^*{}_{\approx 660nm}$ decays much faster than the coherence involving excitation of the low exciton component $(P_{D2}{}^{\delta+}P_{D1}{}^{\delta-})_-{}^*{}_{\approx 675nm}$. This observation can be rationalized as follows: the high exciton component decays via exciton relaxation to the low exciton component and therefore the coherence between the high and the low exciton components, as well as between the high exciton and CT state, decays significantly within 500 fs. Conversely and remarkably, the coherence between the low exciton and the CT state, that is, the reactant and product of the charge separation reaction, undergoes a minor decay. This is consistent with the utilization of *functional* coherence by the PSII RC to drive ultrafast and highly efficient electron transfer from $(P_{D2}{}^{\delta+}P_{D1}{}^{\delta-})_-{}^*{}_{\approx 675nm}$ to $(P_{D2}{}^+P_{D1}{}^-)^{\delta*}{}_{\approx 690nm}$ on the ps time range at cryogenic (80 K) and room temperature (277 K).

**Conclusions**

In conclusion, our results show that the CWT analysis allows us to extract the full potential of the 2DES technique by retaining both the frequency and time information contained in the experimental 2D data by means of *time-resolved* 2D frequency maps. Therefore, the CWT analysis allows a consistent physical interpretation of coherent energy and electron transfer processes and provides an unprecedented detailed view of the charge separation events in photosynthesis: we have shown that the functional vibronic coherence between the electronic states involved in the charge separation process in the PSII RC persists during the time scales of energy and electron transfer allowing the system to perform ultrafast and highly efficient solar to electrochemical energy conversion at both cryogenic and physiologically relevant temperatures.

**Methods**

*Experimental section*. The details of sample preparation, experimental setup and data acquisition for the experimental data presented here can be found in ref. 12. The resolution of the 2D data presented is better than 1 nm in both axes, $\lambda_\tau$ and $\lambda_t$. A detailed description of the 2D data processing procedure is presented in ref. 25.

*Continuous wavelet transform*. The WT analysis refers to several time-frequency transformations and signal processing techniques[31-32] that have in common the utilization of a zero mean and short-time oscillating function $\Psi(T)$ called *mother* wavelet. This function can be adapted to the specific characteristics of the temporal signal in order to optimize the analysis. In the analysis presented in this work we use the complex Morlet wavelet function[42]:

$$\Psi(T) = ((\pi F_b)^{-0.5} \exp(-2i\pi F_c T) * \exp(-T^2/F_b) \qquad \text{Eq. 1}$$

where $F_b$ is a bandwidth parameter and $F_c$ is the wavelet center frequency. The Morlet wavelet is a complex exponential centred at frequency $F_c$, windowed by a zero-mean gaussian function, with $\sigma = (F_b/2)^{0.5}$. These two parameters have to be defined in order to get a good compromise between the frequency and time resolution. A large $F_b$ value (for instance $F_b = 10$) produces high frequency resolution at the expense of low time resolution. On the contrary, a small $F_b$ value ($F_b = 0.5$) produces the opposite effect. Therefore, a range of $F_b$ values has to be explored in order to select the $F_b$ value to obtain optimal frequency and time resolution. In the present work we have chosen $F_b = 2$, and $F_c$ has been fixed to 1. The effect of the $F_b$ in the frequency/time wavelet resolution relation is shown in the Supplemental Information (Fig. S3) for two cross-peaks: (675,681) nm corresponding to the 120 cm$^{-1}$ 2D frequency map and (665,680) nm corresponding to the 340 cm$^{-1}$ 2D frequency map . By comparing the two limiting cases [*top* ($F_b$ = 10) *and bottom* ($F_b$ = 0.5) *panels*] with the optimal case [*middle* ($F_b$ = 2) *panels*], it is clear that increasing excessively the time resolution results in low frequency resolution and vice versa.

The variable *T* in the mother wavelet function is translated by a factor *u* and dilated by a factor *s* called scale, giving the wavelet *atom* function $\Psi^*_{us}(T)$, which provides the effective basis for the transformation:

$$\Psi^*_{u\,s}(T) = \frac{1}{\sqrt{s}} \Psi\left(\frac{T-u}{s}\right) \qquad \text{Eq. 2}$$

In the analysis of the 2D spectra as a function of population time T [2Dspectra(T)] we expand the temporal signal 2Dspectra(T) in the time-frequency domain via the inner product of this input signal with a wavelet atom function $\Psi^*us(T)$, giving rise to the Continuous WT (CWT) as it is shown in Eq. 3:

$$\text{CWF}_{\text{2Dspectra}}(u,s) = \int_{-\infty}^{+\infty} 2\text{Dspectra}(T)\, \Psi^*_{u\,s}(T)\, dT \qquad \text{Eq. 3}$$

The main advantage of WT with respect to the simpler windowed Fourier transform analysis, is that WT realizes a multi-resolution analysis by using different time intervals (scales) for different frequencies. For low frequencies, the WT utilizes long timescales which yields optimal frequency resolution whereas for high frequencies the WT employs short timescales thus achieving better temporal localization. Therefore, the WT is optimally adapted to each of the frequencies involved in the signal. Usually, the CWT is obtained over the entire temporal axis, in this case the population time *T*, by scanning *u* over the full temporal domain of the 2Dspectra(*T*) for a fixed scale *s*. The operation is repeated, when necessary, with a different scale to obtain the CWT for the whole temporal axis and a chosen interval of *s* values.

The wavelet analysis has been performed to each 2D trace independently using the MATLAB[43] function *cwt* (coefs = cwt(x,scales,'*wname*')). This function computes the continuous wavelet transform (CWT) of the real-valued signal x. The wavelet transform is computed for the specified scales using the analyzing wavelet function *wname*. In our study we have used the complex Morlet wavelet (cmor) function as a *mother* wavelet (Eq. 1). This *mother* wavelet function is translated by *u* and dilated by the scale *s*, giving the wavelet *atom* function $\Psi^*_{us}(T)$ (Eq. 1), which provides the effective basis for the transformation. The convolution is computed for a matrix set of the variables *u* and *s*. The matrix has the number of rows equal to the length of scales *s* and number of columns equal to the length of the input signal *u*. The scale variables are fixed by the relation: scales = (1/20fs)*Fc./(frequencies) where 20 fs is the time step of the experimental 2D data analyzed; and the frequencies are the input parameters of the frequencies under study.

**Acknowledgments**

E. R. and R. v. G. were supported by the VU University Amsterdam, the Laserlab-Europe Consortium, the TOP grant (700.58.305) from the Foundation of Chemical Sciences part of NWO and the advanced investigator grant (267333, PHOTPROT) from the European Research Council. E. R., M. B. P and R. v. G. were supported by the EU FP7 project PAPETS (GA 323901). R.v.G. gratefully acknowledges his Academy Professorship from the Netherlands Royal Academy of Sciences (KNAW).V. I. N. was supported by Russian Foundation for Basic Research (grant No. 15-04-02136) and by a NWO visitor grant. J.P. was supported by Ministerío de Economía y Competitividad Project No. FIS2015-69512-R and the Fundación Séneca Project No. 19882/GERM/15. S. E. M. and A. W. C. were supported by the Winton Programme for the Physics of Sustainability. S. E. M. was supported by EPSRC. M. B. P. was supported by an ERC Synergy Grant (319130, BioQ).

**Author contributions statement**

J. P. developed the wavelet data analysis program and performed the analysis. E. R., J. P., A. W. C. and R. v. G. discussed and interpreted the analyzed data. E. R. performed the windowed Fourier transform analysis. A. W. C. performed the simulation of the 2D spectral dynamics. S. E. M. performed the analysis entitled "Investigating the origins of the amplitude modulations observed in the wavelet data" presented in the Supplementary Information. E. R., J. P., A. W. C., S. E. M. and R. v. G wrote the paper. All authors discussed the results and commented on the manuscript.

**Competing financial interests statement**

The authors declare no competing financial interests.